# Monte Carlo simulation of spin polarized transport in nanowires and 2-D channels of III-V semiconductors

Swetali Nimje*, Ashutosh Sharma and Bahniman Ghosh

*Department of Electrical Engineering, Indian Institute of Technology, Kanpur 208016, India*

**Abstract**

We simulated spin polarized transport of electrons along III-V nanowires and two dimensional III-V channels using semi classical Monte Carlo method. Properties of spin relaxation length have been investigated in different III–V zinc-blende materials at various conditions, such as, temperature, external field etc. Spin dephasing in III-V channels is caused due to D'yakonov-Perel (DP) relaxation and due to Elliott-Yafet (EY) relaxation. Spin dephasing length in nanowire is found to be greater than that in 2-D channel.

*Email : swetali@iitk.ac.in , nimje.swetali2205@gmail.com

## Introduction

Spintronics [1-5], which exploits both the property of electric charge and spin, is the emerging field in last few decades. Spintronics based devices require control of the spin polarization in the device channel. The spin field effect transistor (Spin-FET) developed by Datta and Das [6] which uses ferromagnetic material as the source and drain is an example of spintronics based devices. Devices using the spin related properties of semiconductors have been extensively investigated [6-11]. These devices combine storage, detection, logic and communication capabilities to produce a multifunctional device on a single chip which could replace several components.

For improving these devices, many challenges have to be overcome which include finding better methods of polarizing a spin system controlling the duration for which the system is able to memorize its spin orientation and efficient detection of spin. These represent some of the fundamental issues associated with carrier's spin apart from its charge related challenges. Based on them, we can narrow down to three main processes which form the basis of the functioning of any spintronics device, which are spin injection at the source, spin transport through the material and spin detection at the drain. In this paper, we focus on the study of spin transport [12]. Spin relaxation is a basic process in the study of spin transport. Spin dephasing length is an important property apart from spin dephasing time as shown in research [12, 13].

Rigorous theoretical and experimental studies have been done to study spin transport in semiconductors and metals. III-V semiconductor materials have a nice blend of remarkable properties like high carrier mobility and high saturation velocity and their band structures are well suited for optical applications and spintronics devices and this distinguishes them from other materials for research .Thus thorough research has been done to study their spin properties [14-17]. Spin transport in GaAs 1-D channel is studied at the liquid nitrogen temperature at different values of driving electric fields [18] which can be developed to design a gate controlled spin interferometer where the suppression of spin dephasing is a vital issue.

In this paper we have studied the spin dephasing in 1D III-V nanowire and 2D III-V channels at various conditions like electric field and temperature .We have reported the simulation results for electric field in the range of 75V/cm- 2kV/cm and temperature in the range of 77K-323K for 1-D III-V channel and 2-D III-V channel and compared them. Various III-V materials for which study is done are InAs, GaSb, GaAs and GaP.

## Model

A detailed explanation of the Monte Carlo method and spin transport model is given in the references [18-22]. In this paper we consider the relevant features of the model. According to the co-ordinate system chosen, x is along the length of the device, y is along the width of the device and z is the along the thickness of the device. In the 2-D system the electrons are restricted in the z- direction, while in the 1-D system electrons are restricted in the y direction and the z direction.

III-V compounds have bulk inversion asymmetry which causes Dresselhaus spin-orbit interaction [23].The transverse field acts as a foremost symmetry breaking electric field which breaks the structural inversion asymmetry and this leads to Rashba spin orbit coupling [24]. Spin-orbit Hamiltonian is influenced by the electron spin which comprises of the Dresselhaus interaction and Rashba interaction. For nanowire Dresselhaus spin–orbit Hamiltonian [23] is expressed as

$$H_D^{1D} = -\beta\left(< k_y >^2 - < k_z >^2\right) k_x\, \sigma_x \quad (1)$$

and for 2-D channel Dresselhaus spin–orbit Hamiltonian is given by,

$$H_D^{2D} = \beta(< k_z >^2)(k_y \sigma_y - k_x\, \sigma_x) \quad (2)$$

For a nanowire Rashba spin orbit Hamiltonian is expressed as,

$$H_R^{1D} = -\eta k_x\, \sigma_y \quad (3)$$

and for a 2-D channel Rashba spin –orbit Hamiltonian is expressed as,

$$H_R^{2D} = \eta\, ( k_y \sigma_x - k_x\, \sigma_y ) \quad (4)$$

The constants β and η depend on the material. η also depends on the external transverse electric field, the expression of η [25] is given below,

$$\eta = \frac{h^2}{2m^*}\frac{\Delta}{E_g}\frac{(2E_g+\Delta)}{(E_g+\Delta)(3E_g+2\Delta)}eE \tag{5}$$

where Δ is the spin orbit splitting of the valence band, $e$ is the electronic charge, $m^*$ is the effective mass, $E_g$ is the band gap and $E$ is the transverse electric field.

The temporal evolution of the spin vector during the free flight time is given by following equation [22, 26]

$$\frac{d\vec{S}}{dt} = \vec{\Omega} \times \vec{S} \tag{6}$$

Precession vector ($\vec{\Omega}$) consists of two components, one from the Dresselhaus interaction and other from the Rashba interaction and can be written as [22, 26],

$$\Omega_D = -\frac{2\beta_{eff}k_x\,\hat{\imath}}{\hbar} \tag{7}$$

$$\Omega_R = -\frac{2\eta k_x\hat{\jmath}}{\hbar} \tag{8}$$

Where $\beta_{eff} = \beta(<k_y>^2 - <k_z>^2)$

Using Eq. (7) and Eq. (8) in Eq. (1) and expressing spin vector as $\vec{S} = \overrightarrow{S_x\hat{\imath}} + \overrightarrow{S_y\hat{\jmath}} + \overrightarrow{S_z\hat{k}}$, we get the following relations for each component of spin,

$$\frac{dS_x}{dt} = -\frac{2}{\hbar}\,\eta k_x S_z \tag{9}$$

$$\frac{dS_y}{dt} = -\frac{2}{\hbar}\,\beta_{eff}k_x S_z \tag{10}$$

$$\frac{dS_z}{dt} = -\frac{2}{\hbar}\,k_x(\eta S_x - \beta_{eff}S_y) \tag{11}$$

The whole simulation time is divided into small time steps Δt and the spin components are updated after every Δt. Due to the presence of driving electric field and scattering processes, electron wave vector changes and produces a distribution of momentum states. These in turn cause distribution of spin states resulting in ensemble dephasing. This is the D'yakonov-Perel (DP) relaxation [27]. There is another type of dephasing mechanism, Elliott-Yafet (EY) [28] relaxation which causes instantaneous spin flip known as a spin-flip scattering. The spin relaxation time is given by, [29]

$$\frac{1}{\tau_s^{EY}} = A\left(\frac{k_bT}{E_g}\right)^2 \alpha^2\left(\frac{1-\alpha/2}{1-\alpha/3}\right)\frac{1}{\tau_p} \tag{12}$$

where $E_g$ is the band gap, α = Δ/ ($E_g$ + Δ), where Δ is the spin orbit splitting and $\tau_p$ is the momentum relaxation time. A is a dimensionless constant and varies between 2 and 6. We have used A = 4.

In III-V compounds the conduction band [30] is described by Γ-valley along with L-valley and X-valley. L-valley and X-valley possess higher energy than the Γ-valley. We have only taken lowermost Γ-valley in our simulation and assumed that other two are higher up in the energy level and hence depopulated.

According to Conwell and Vassal [20], non- parabolicity approximation of the band is considered by following energy- wave vector relation:

$$\epsilon(1+\alpha\epsilon) = \frac{\hbar^2k^2}{2m} \tag{13}$$

Where α (non parabolicity parameter) is given by the following expression,

$$\alpha = \frac{1}{E_g}\left(1-\frac{m}{m_0}\right)^2 \tag{14}$$

In our simulation ,the scattering rates for 1-D and 2-D channel considered are optical phonon scattering[31,32], acoustic phonon scattering[31,32], polar optical phonon scattering[32,33] and surface roughness scattering[31,34] .

## Results

We have performed the semi classical Monte Carlo simulation described in previous section to simulate spin polarized electron transport in nanowires and along the 2-D channel made of III-V semiconductor materials. The nanowire consists of 5nm x 5nm cross section whereas 2-D channel consists of 5nm x 100nm cross-section. The transverse effective field is taken to be 100kV/cm [15] which results in Rashba spin orbit coupling. In simulation four subbands [30] are taken in both channels. Moderate values of driving electric field, in range of 75V/cm-2kV/cm, are taken such that majority of electrons are confined to first four bands. The subbands' energy levels are considered using infinite potential well approximations. Also the transverse dimension of the channel is considered to be very small (5nm) due to which higher subbands will have a very high energy level and thus can be considered as depopulated.
The material parameters used for Monte Carlo simulation are given in the table 1 [35-41].

| Parameter | InAs | GaSb | GaAs | GaP |
|---|---|---|---|---|
| Effective mass | $0.023m_o$ | $0.041m_o$ | $0.063\ m_o$ | $0.09m_o$ |
| Bandgap (eV) at 300K | 0.354 | 0.726 | 1.424 | 2.26 |
| Density (g/cm$^3$) | 5.68 | 5.61 | 5.32 | 4.14 |
| Speed of sound(cm/s) | $3.8\ x10^5$ | $6.07x10^5$ | $5.5x10^5$ | $5.83x10^5$ |
| Static dielectric constant ($\varepsilon_S$) | 15.15 | 15.7 | 12.9 | 11.1 |
| acoustic phonon deformation potential (eV) | 4.9 | 8.9 | 8.8 | 7.14 |
| Polar optical phonon energy (eV) | 0.030 | 0.0297 | 0.035 | 0.051 |
| Spin Orbit Splitting (eV) | 0.41 | 0.80 | 0.34 | 0.08 |

**Table.1.** Material parameters used for simulation

The simple theoretical formula [42] used for calculation of Lange –g factor is given by

$$g = 2 - \frac{2}{3}\frac{Ep\,\Delta}{(Eg+\Delta)Eg}$$

where Eg is the energy band gap, Δ is the spin-orbit splitting energy, and Ep is the energy equivalent of the principal interband momentum matrix element. Ep is always taken to be 22 eV.

A step size of Δt=0.2 fs in time is selected and the simulation is run for 5 x $10^5$ such time steps which allows the electrons to achieve steady state. Data are recorded for the final 30,000 steps only and ensemble average is calculated for each component of the spin vector for the last 30,000 steps at each point of the wire according to the following expression [24],

$$< S_i > (x,t) = \frac{\sum_{t=t1}^{t=T}\sum_{n=1}^{n_x(x,t)} s_{n,i}(t)}{\sum_{t=t1}^{t=T} n_x(x,t)}$$

where i represents the x, y and z components, $n_x(x,t)$ is the total number of electrons in a grid of distance Δx around position x at time t, $s_{n,i}(t)$represents the value of the i$^{th}$ spin component of the n$^{th}$ electron at time t. Here T is the end time and t1 is the time at which we start recording the data. The magnitude of the average spin vector is then calculated using the expression

$$|< S > (x,T)| = \sqrt{< S_x >^2 + < S_y >^2 + < S_z >^2}$$

Spin dephasing length is defined as the distance from the source (x=0, where the electrons are introduced) where |<S>| falls to 1/e times of its initial value at injection. The electrons are injected with an initial polarization of 1 and therefore the initial value of |<S>| is 1.

*Spin dephasing length at the room temperature (300K) and for driving electric field of 1kV/cm*

Figures 1-4 show the magnitude of ensemble averaged spin of various III-V compounds along nanowire and 2-D channel at room temperature (300K) for driving electric field of 1kV/cm. Spin dephasing lengths for various III-V compounds along 1-D and 2-D channel are given in Table 2.

| III-V compound | 1-D Channel | 2-D Channel |
|---|---|---|
| GaAs | 26.56μm | 1.592μm |
| GaP | 27.12μm | 1.584μm |
| GaSb | 11.76μm | 0.504μm |
| InAs | 10.40μm | 0.392μm |

**Table.2.** Spin Dephasing lengths for various III-V compounds at 300K for driving electric field of 1kV/cm

The spin dephasing length in nanowire is observed to be larger than that of 2-D channel. References [18, 43-45] report similar results which show the improvement in spin dephasing length in 1-D channel over 2-D. The difference in length is due to the dominance of different spin dephasing mechanism in these channels. DP relaxation is suppressed in nanowire [43, 44]. Thus a nanowire has significantly lesser DP relaxation than a 2-D channel and causes larger dephasing lengths. The comprehensive justification of this effect is accounted for in reference [30]. Spin dephasing increases with the randomness of the motion of electron. In a 2-D channel, the electron motion arbitrarily occurs along two directions whereas in a nanowire it occurs only in one direction.

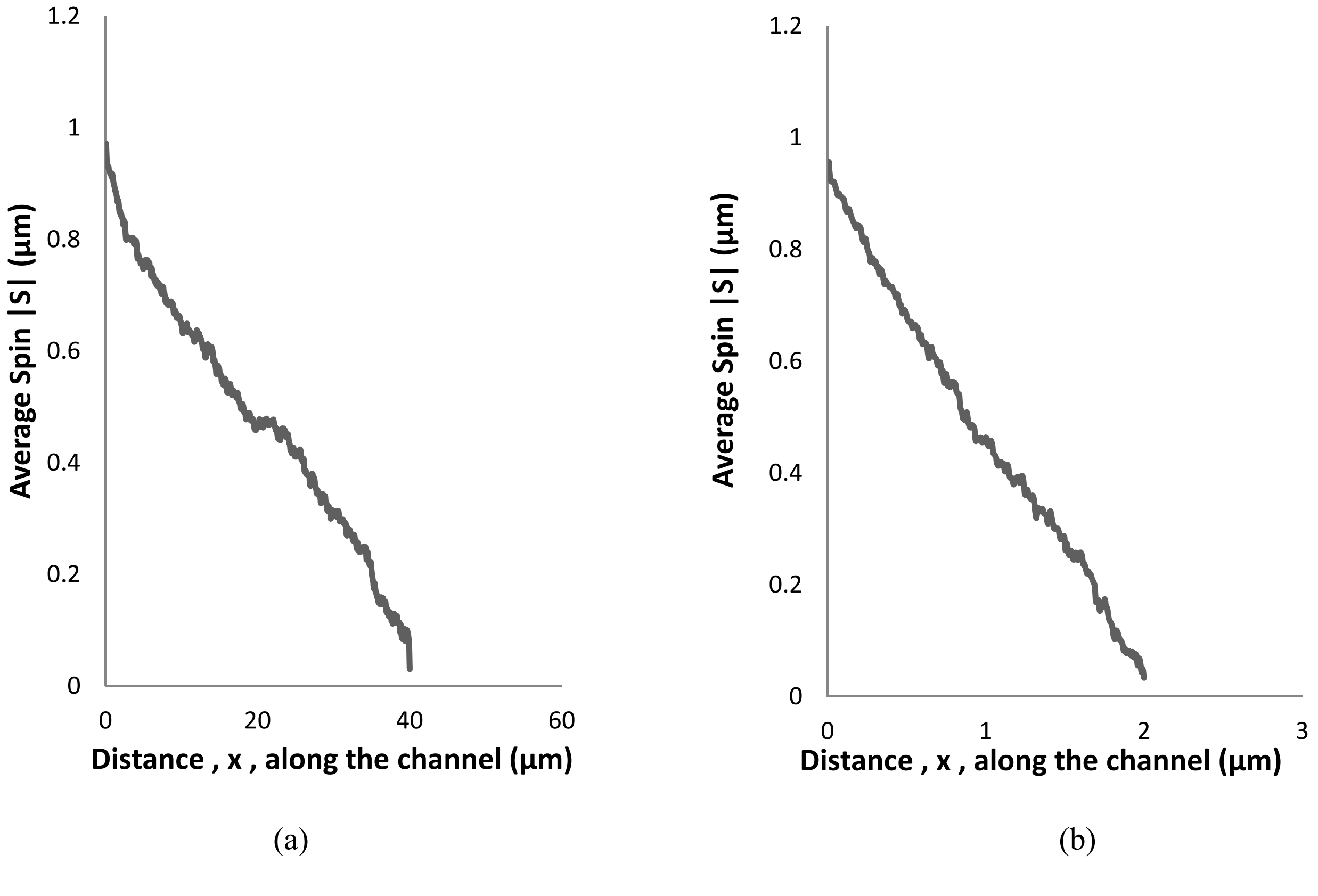

Fig.1. Spin decay along a GaAs (a) nanowire (b) 2-D channel for injection polarization along the z-direction at 300K at driving electric field of 1kV/cm.

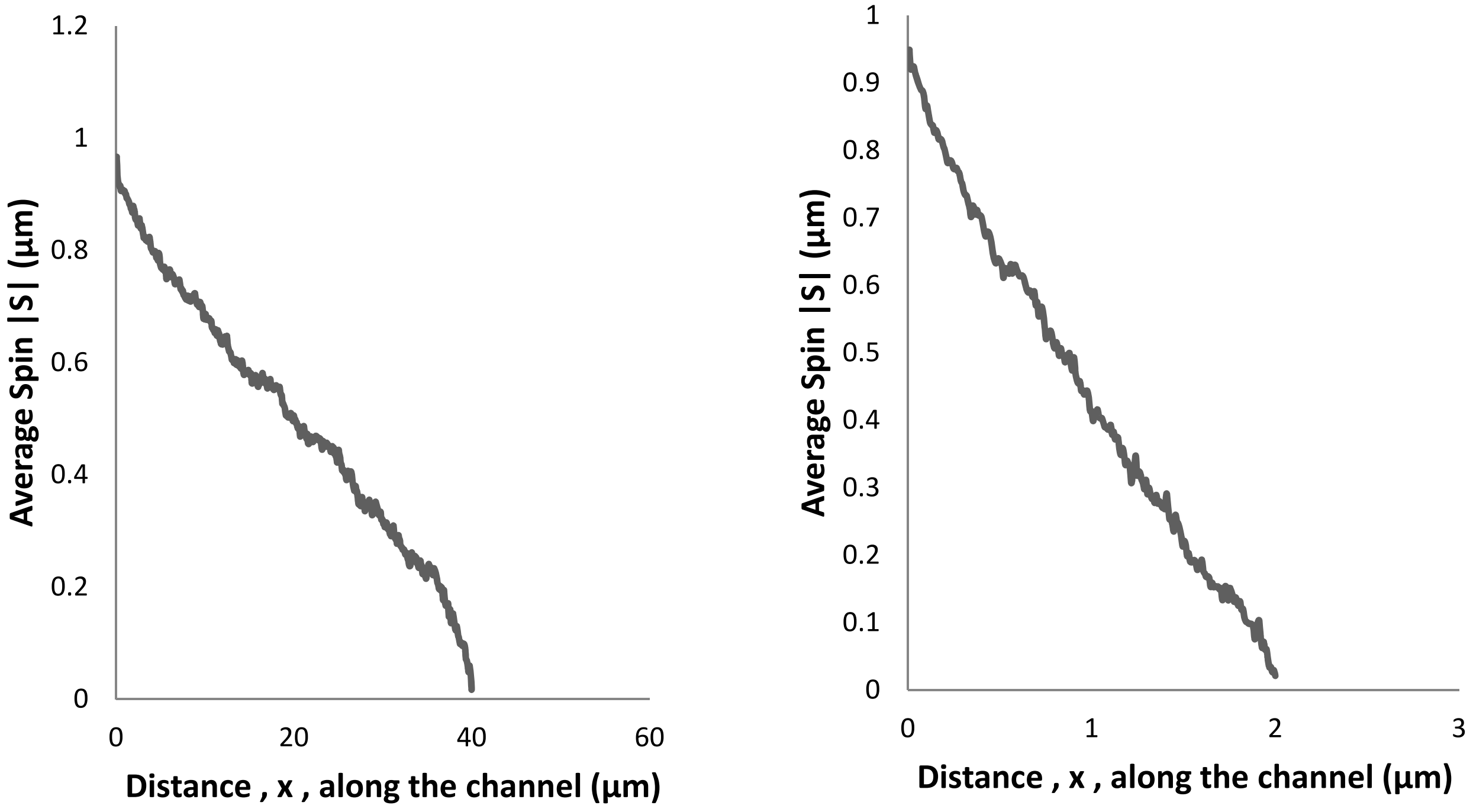

(a) (b)

Fig.2. Spin decay along a GaP (a) nanowire (b) 2-D channel for injection polarization along the z-direction at 300K for a driving electric field of 1kV/cm.

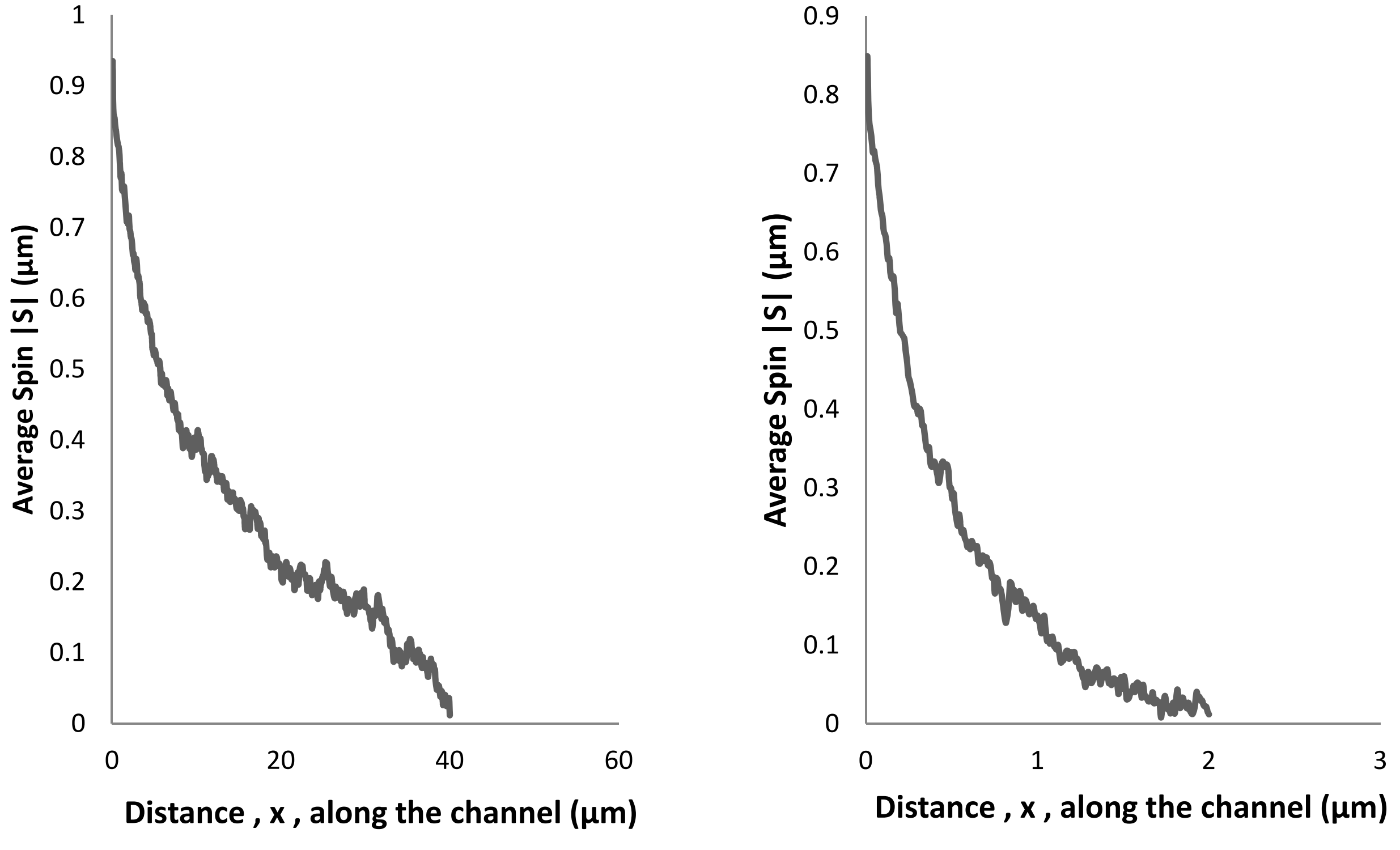

(a) (b)

Fig.3. Spin decay along a GaSb (a) nanowire (b) 2-D channel for injection polarization along the z-direction at 300K a for driving electric field of 1kV/cm

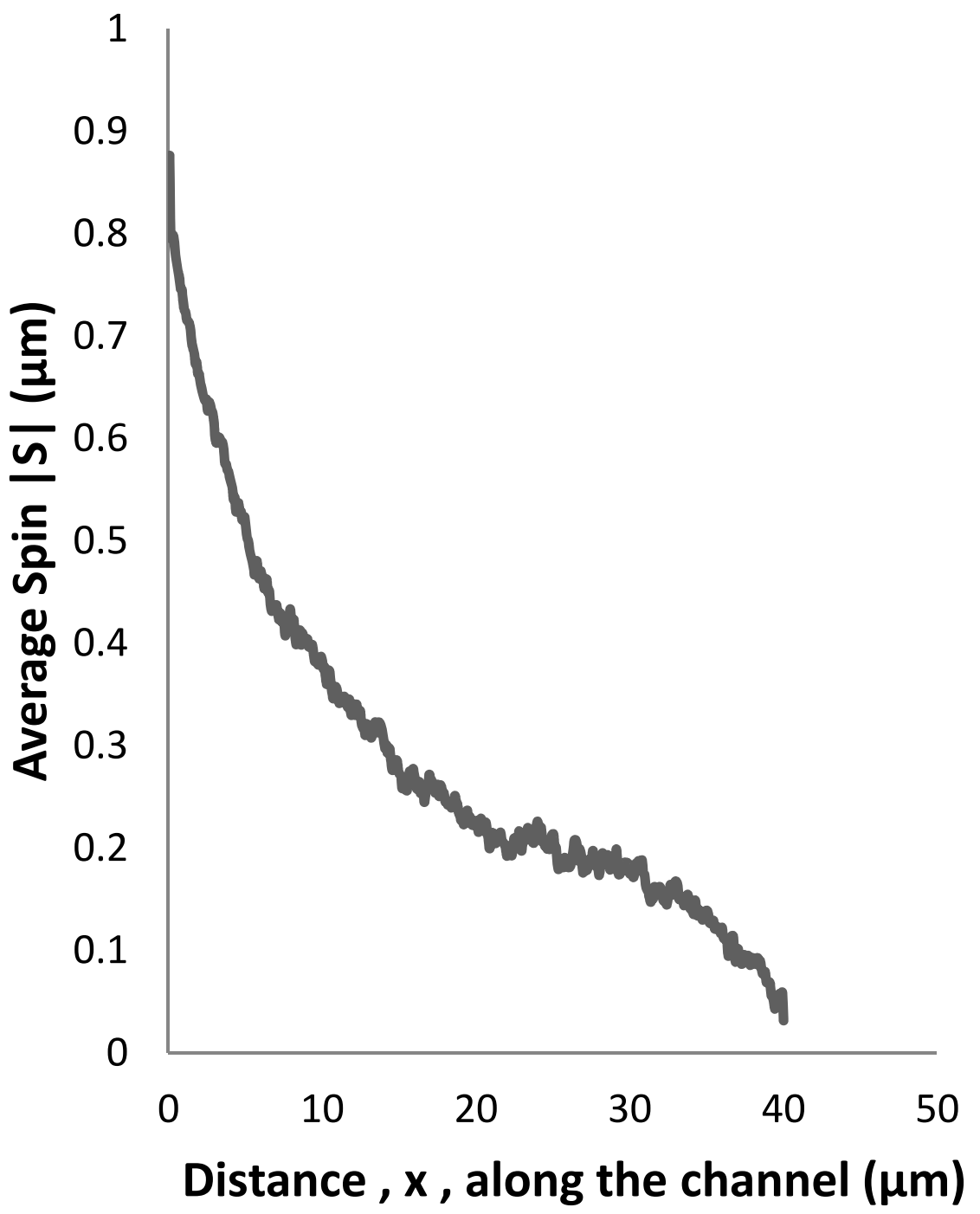

(a)

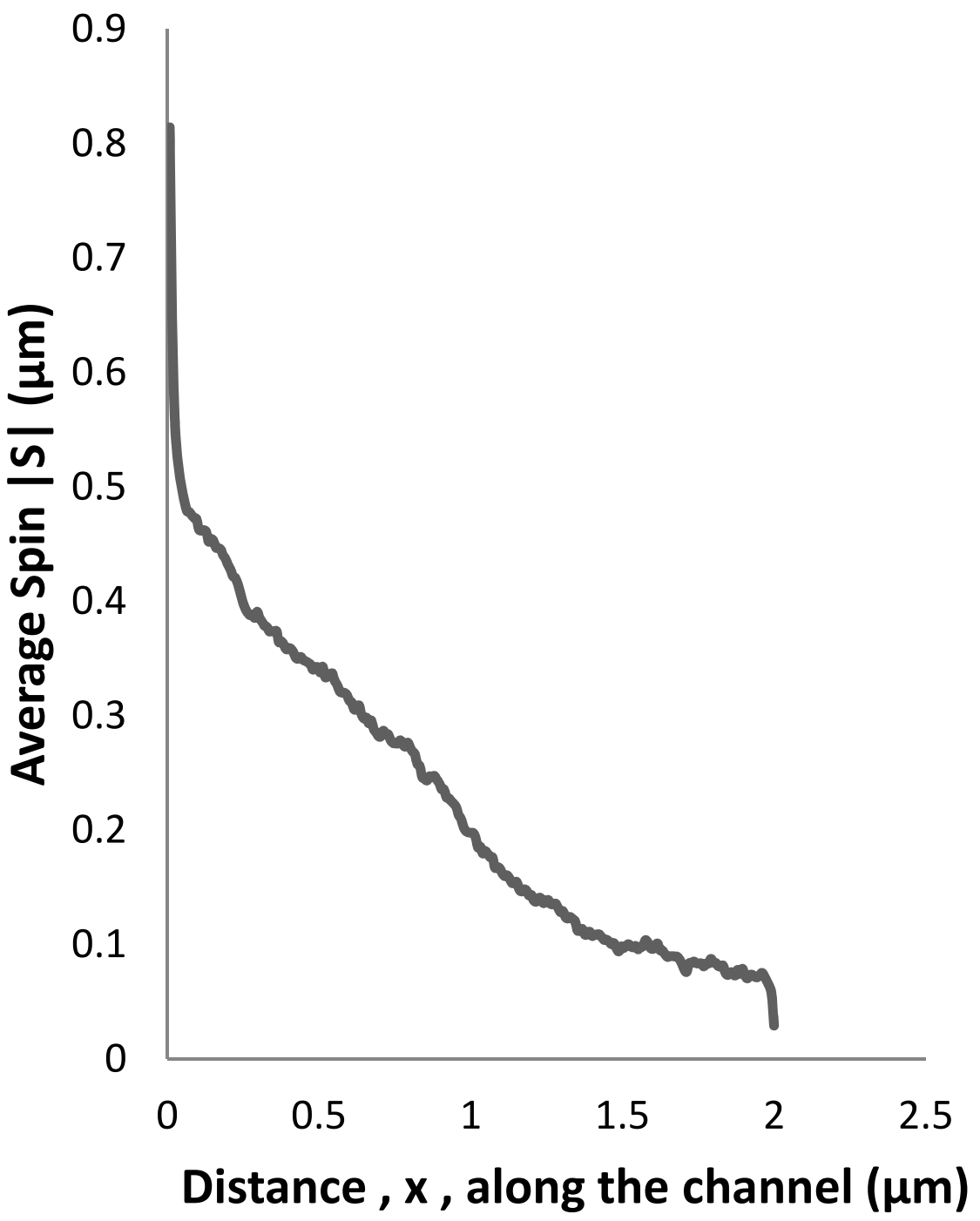

(b)

Fig.4. Spin Decay along a InAs (a) nanowire (b) 2-D channel for injection polarization along the z-direction at 300K for a driving electric field of 1kV/cm

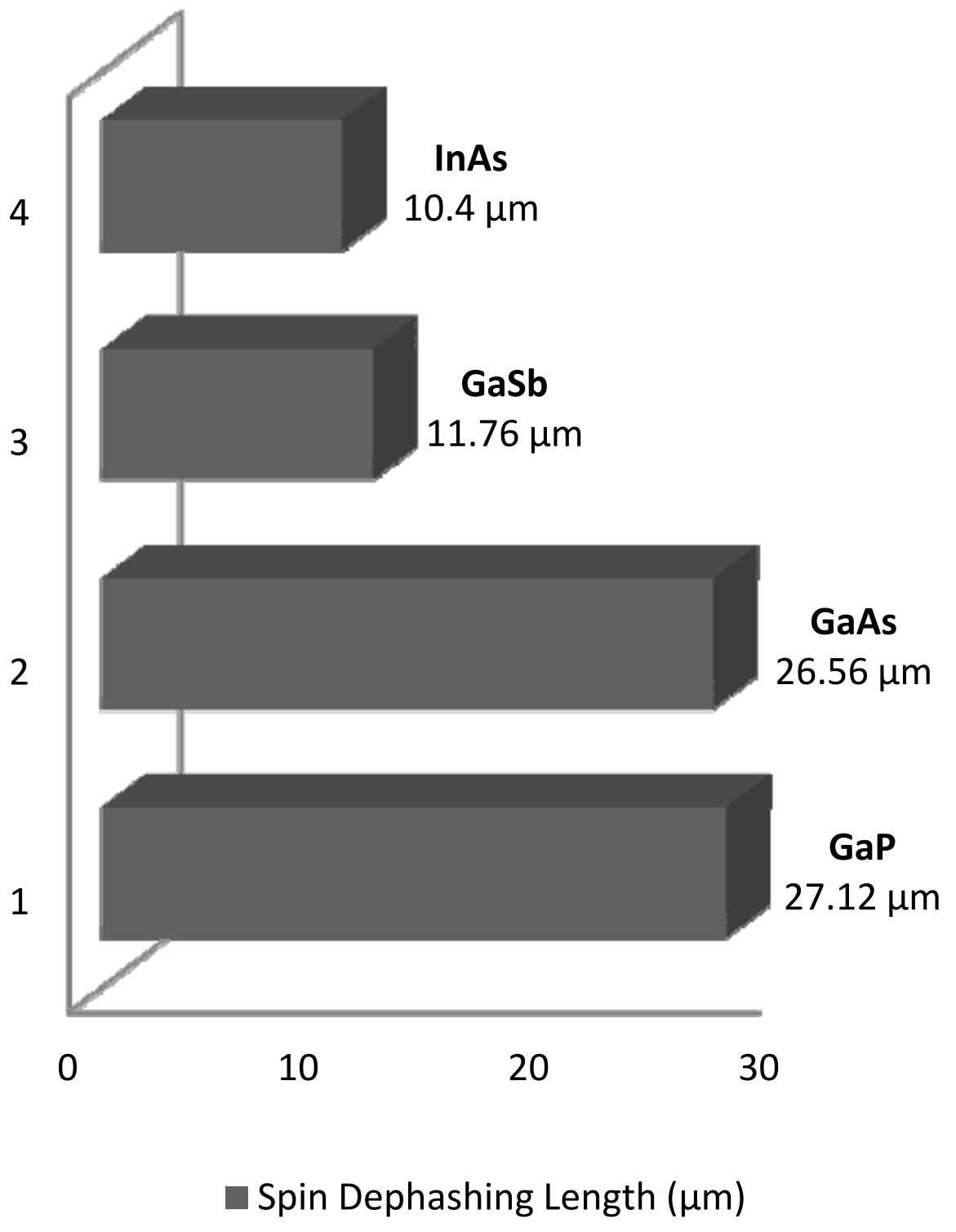

(a)

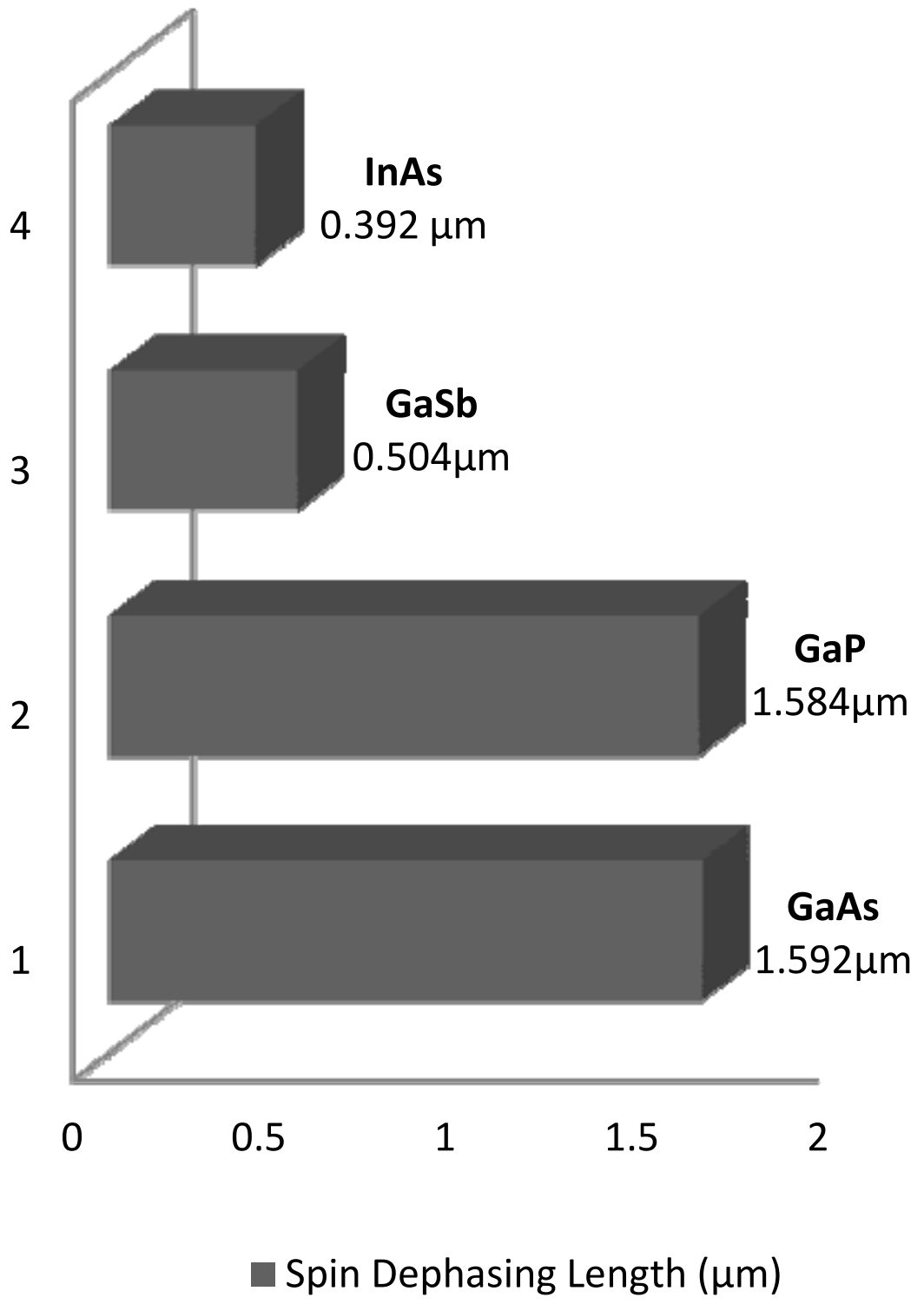

(b)

Fig.5.Bar graph showing the comparison between spin dephasing lengths among various III-V semiconductors for (a) 1-D Channel (b) 2-D Channel.

Figure 5 (a) shows that for both 1-D and 2-D channels, GaAs and GaP have only a slight difference in their spin dephasing lengths. Similarly, GaSb and InAs also have comparable value of spin dephasing length. There is a large difference between the spin dephasing lengths of GaAs and InAs and between GaP and GaSb. This difference arises due to the difference in depolarization rates because of DP relaxation and EY mechanisms. The spin orbit coupling in GaSb and InAs is much larger than in GaAs and GaP. The value of Rashba coefficient at 300K and at the transverse electric field 100kV/cm from equation (3) for III-V material is given in Table 3.

| III-V compound | GaAs | GaP | GaSb | InAs |
|---|---|---|---|---|
| Rashba Coefficient (η) | $2.77\times10^{-32}$ | $0.67\times10^{-32}$ | $63.51\times10^{-32}$ | $240.83\times10^{-32}$ |

**Table.3.** Rashba Coefficient at 300K at the transverse electric field of 100kV

Thus the Rashba spin orbit interaction is stronger in GaSb and InAs. This leads to stronger DP relaxation and thus faster dephasing in them as compared to GaAs and GaP. Also InAs and GaSb are narrow gap semiconductors with very high spin orbit coupling (Table 1) whereas GaAs and GaP are wide bandgap semiconductor with weak spin orbit coupling (Table.1). Therefore the Elliott Yafet spin relaxation mechanism is strongly dominant which results in faster depolarization in InAs and GaSb. Thus the spin dephasing lengths are longer in GaAs and GaP compared to GaSb and InAs.

*Effect of applied electric field*

Figures 6-9 show the variation of spin dephasing length with driving electric field. As seen from the figures, spin dephasing length is weakly dependent on electric field showing nonmonotonic behavior because of two opposing features: scattering rates and ensemble average drift velocity.

When drift velocity is prevailing over the scattering rates, the spin penetrates further into the channel resulting in larger spin relaxation lengths. However, when scattering rates dominate drift velocity, they dephase the spin faster. The overall effect is determined by the dominating effect from amongst the scattering rate and drift velocity.

At higher electric field, spin dephasing length increases for both 1-D and 2-D channels with increase in driving electric field. Since in our model, we have considered four subbands for which the scattering rates saturate at higher field value. It remains almost constant and doesn't rise after a certain value of electric field. Values of electric field are chosen such that drift velocity doesn't get saturated. Thus in high electric field region, drift velocity dominates over scattering rates. This results in increased spin dephasing length.

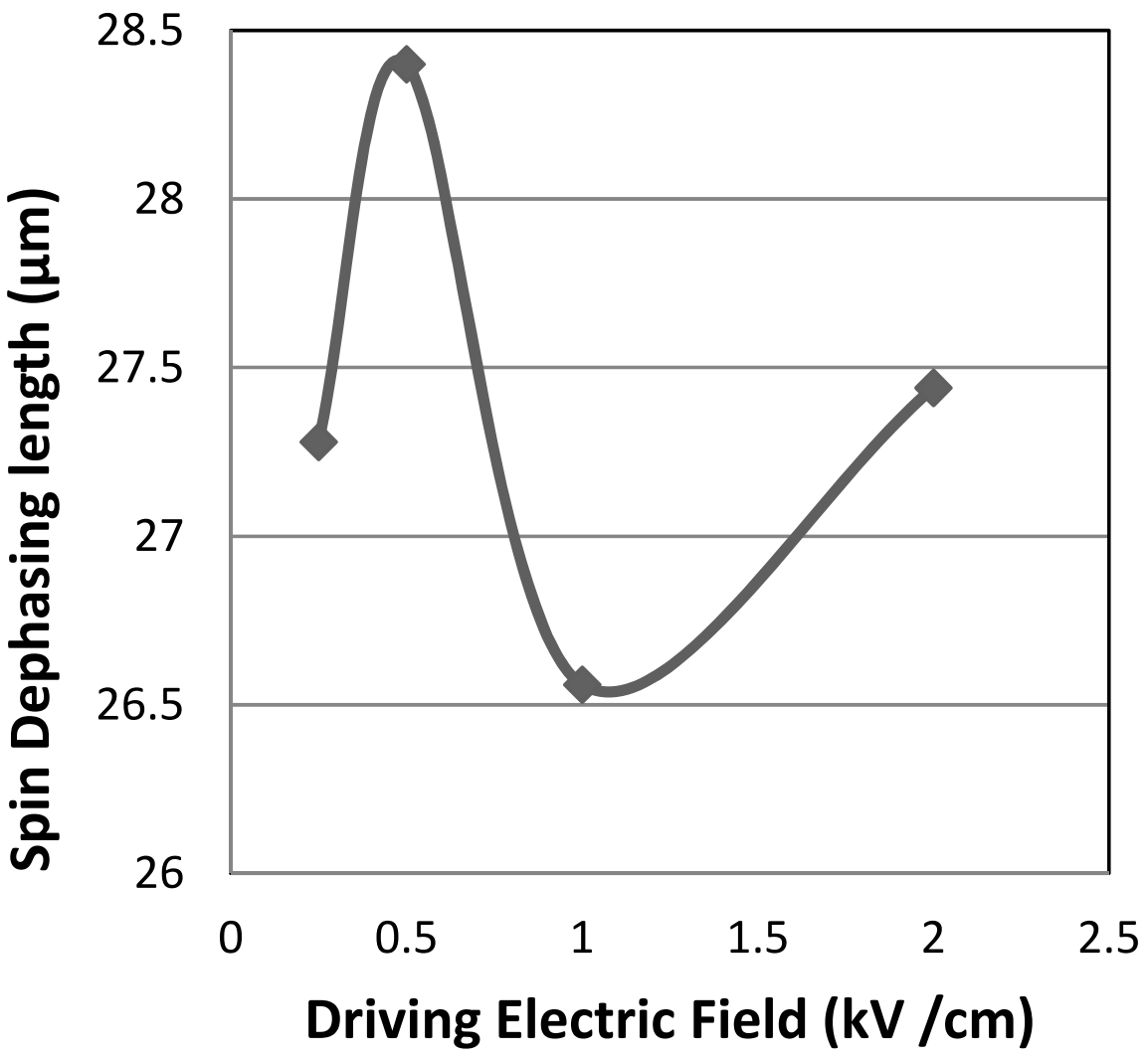

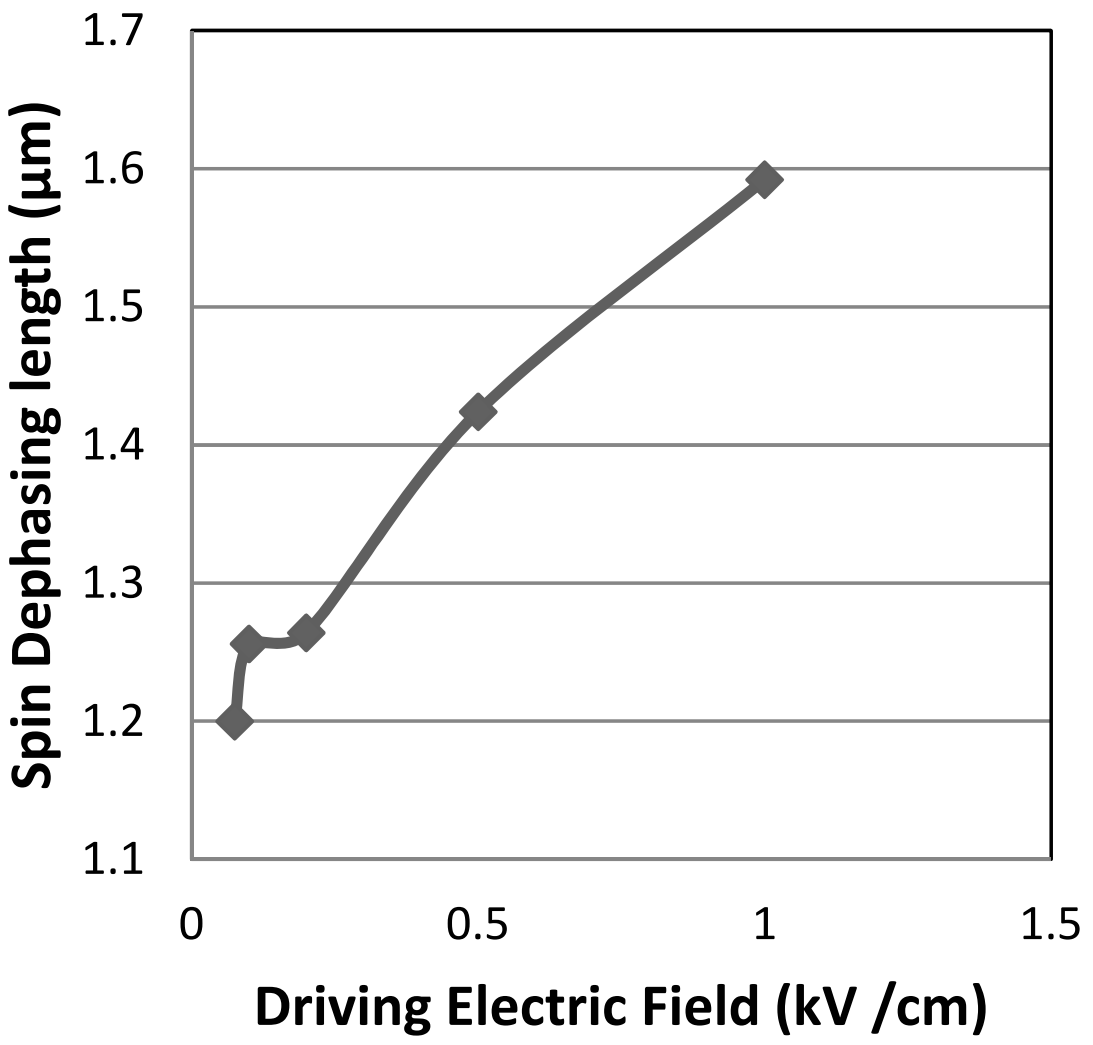

(a) (b)

Fig.6. Variation of spin dephasing length with driving electric field for GaAs (a) nanowire (b) 2-D channel

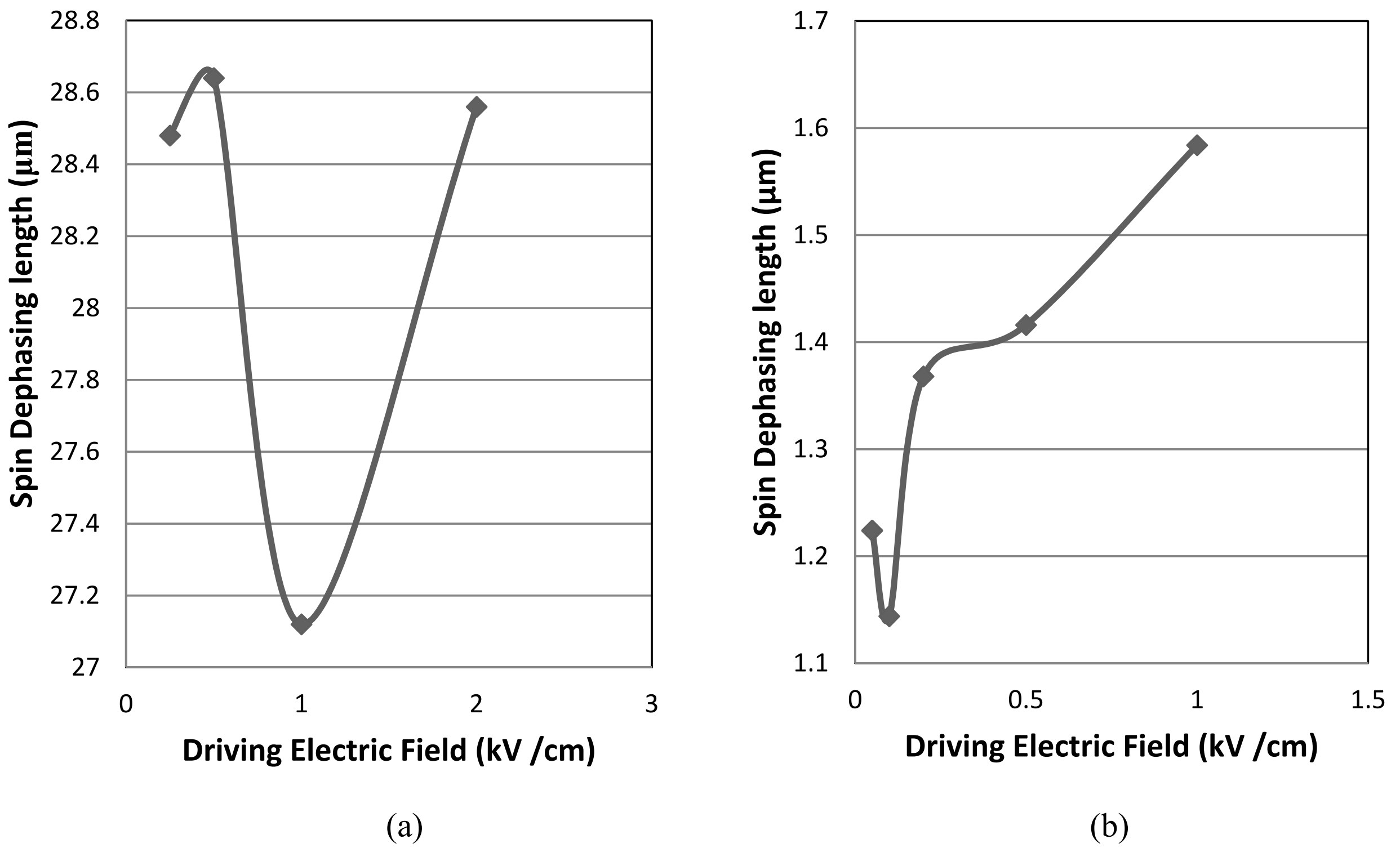

Fig.7. Variation of spin dephasing length with driving electric field for GaP (a) nanowire (b) 2-D channel

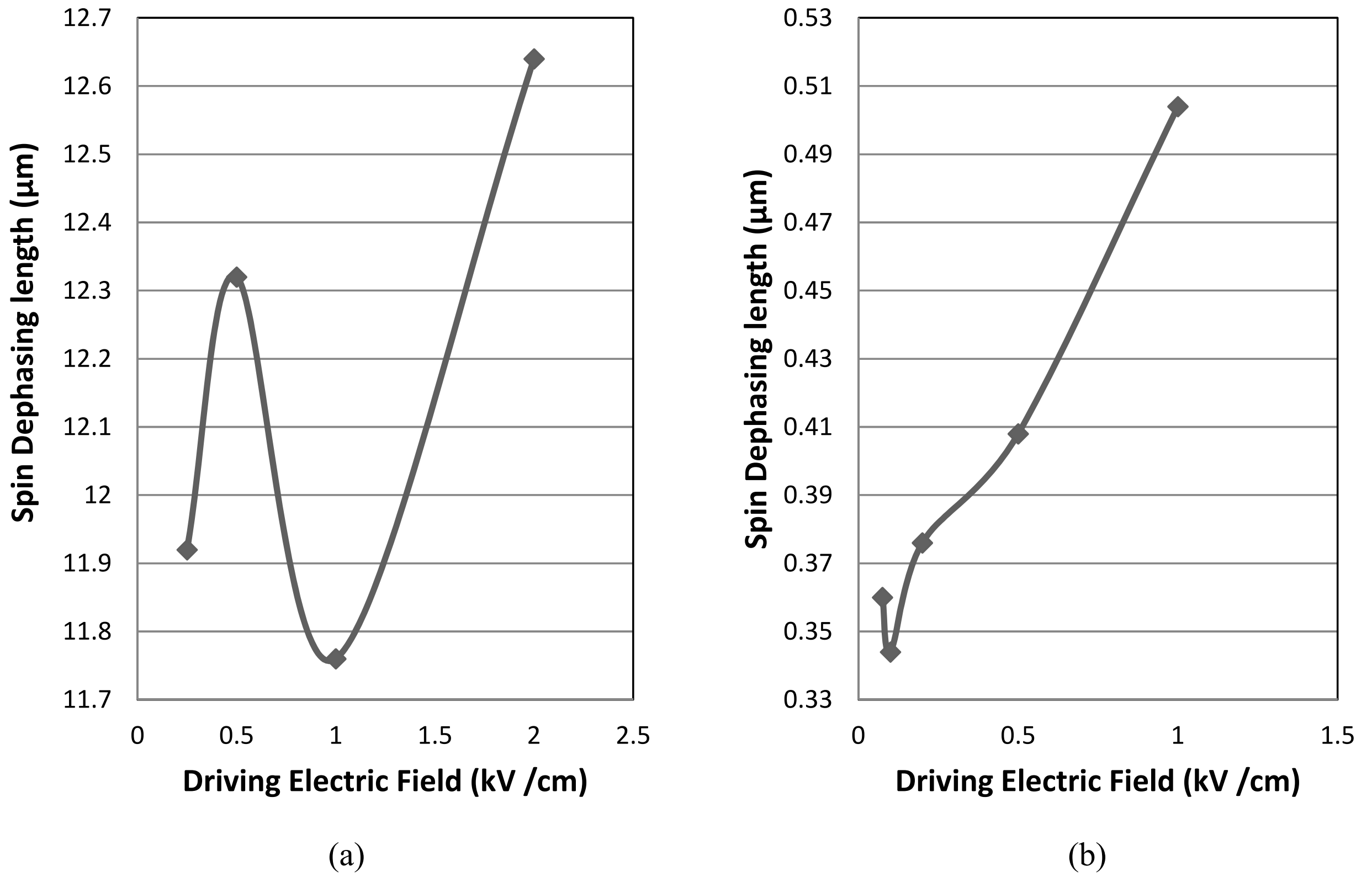

Fig.8. Variation of spin dephasing length with driving electric field for GaSb (a) nanowire (b) 2-D channel

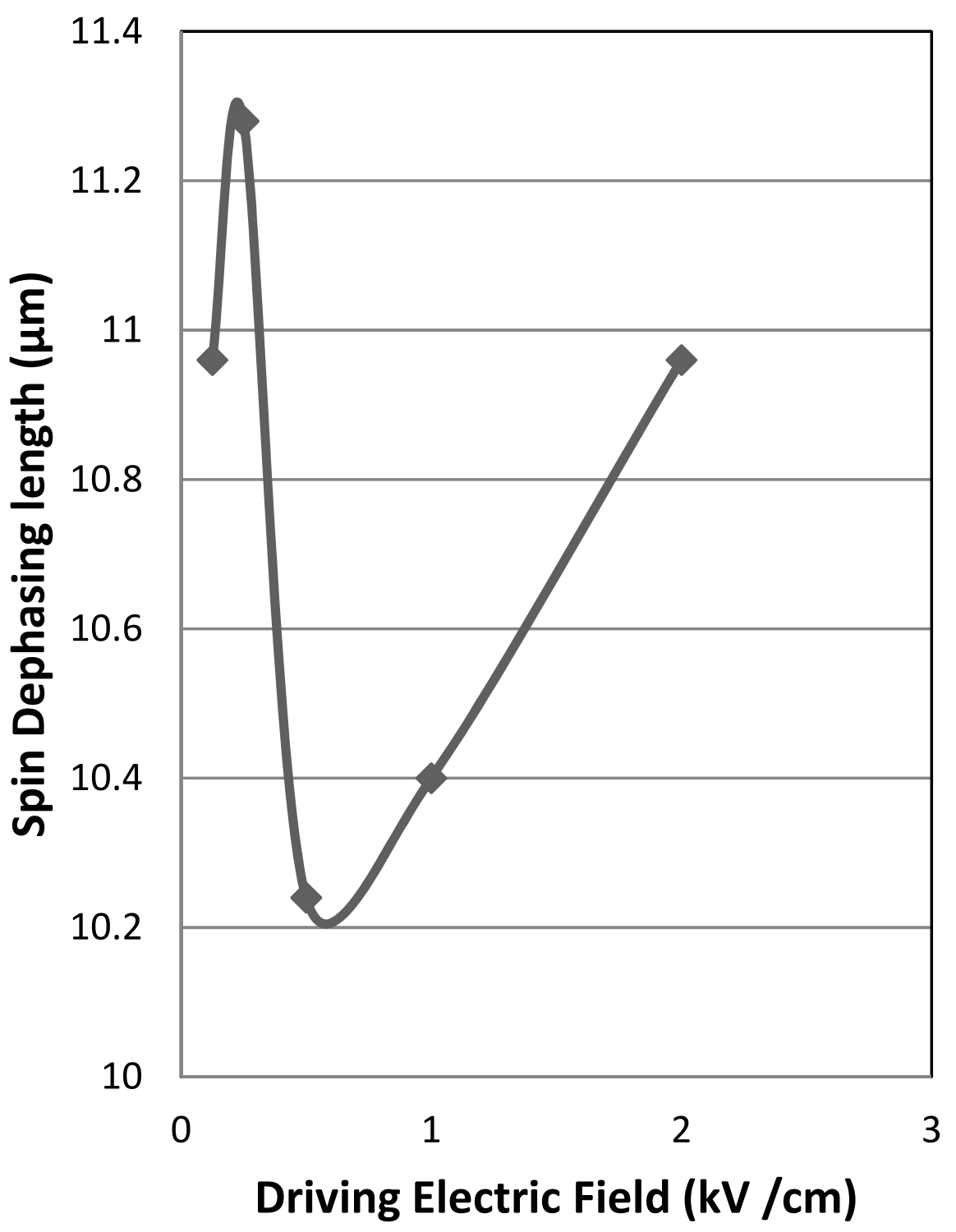

(a)

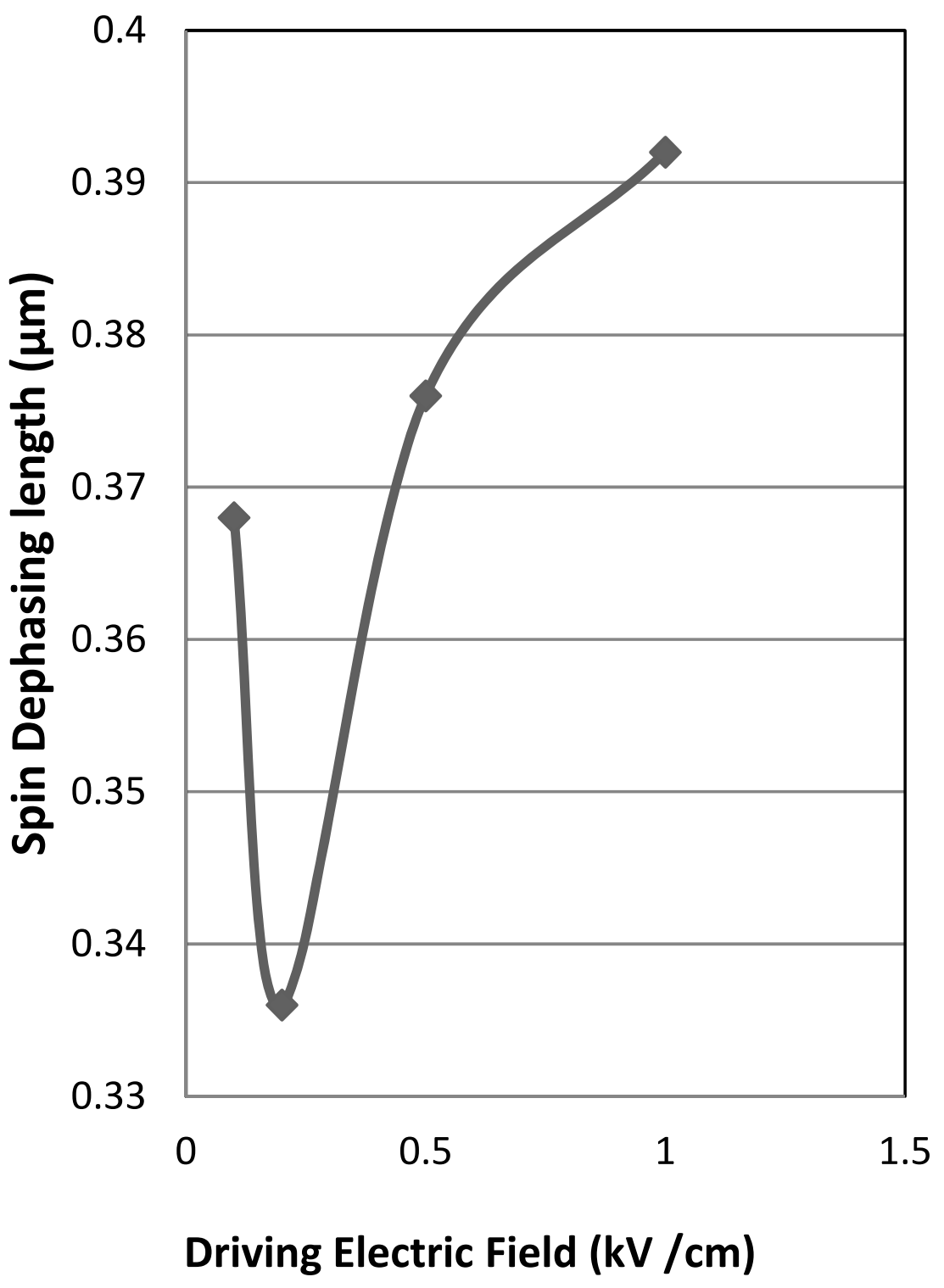

(b)

Fig.9. Variation of spin dephasing length with driving electric field for a InAs (a) nanowire (b) 2-D channel

*Effect of temperature*

Figures 10-13 show the variation of Spin dephasing length with temperature. It is inferred from the graphs that when temperature is increasing the spin dephasing length is decreasing.
Increasing temperature leads to increase in acoustic phonon scattering for both 1-D and 2-D channel which causes randomization of $k$ and $\Omega's$ [26]. This leads to faster dephasing, thus the dephasing length decreases with increase in temperature. Temperature dependence can be clearly seen from the expression of scattering rate for 1-D channel [31]

$$\Gamma_{nm}^{ac}(k_x) = \frac{\mathrm{E}_{ac}{}^2\ \mathrm{k}_B T\ \sqrt{2m^*}}{\hbar^2\ \rho v^2}\ D_{nm} \frac{(1+2\alpha\varepsilon_f)}{\sqrt{\varepsilon_f(1+\alpha\varepsilon_f)}}(\varepsilon_f)$$

where $\mathrm{E}_{ac}$ is the acoustic deformation potential, $\rho$ is the crystal density, $v$ is the sound velocity and $\Theta$ is the Heaviside step-function. $D_{nm}$ is the overlap integral [31] associated with the electron-phonon interaction and that of 2-D channel [32]

$$S_{acp,i}(k) = \frac{D_i^2 k_B T m_i^*}{\hbar^3 s_1^2} \int \left|F_{i,m}(z)^2\right|\left|F_{i,n}(z)^2\right| dz\ \times \left[1 + 2\alpha_i \varepsilon_{i,m}(k)\right] \cdot u[\varepsilon_{i,m}(k) - \varepsilon_{i,n}(0)]$$

where $D_i$ is the acoustic deformation potential in the $i^{th}$ valley, $\rho$ is the density of the material, $s_l$ is the sound velocity in the material and $u(x)$ is the step function. $F_{i,m}(z)$ is the electron wavefunction of the $m^{th}$ subband in valley $i$. $\varepsilon_{i,m}(k)$ is the energy of the electron in the $m^{th}$ subband of the $i^{th}$ valley with the wavevector k. $\varepsilon_{i,n}(0)$ is the $n^{th}$ subband energy level in the $i^{th}$ valley.

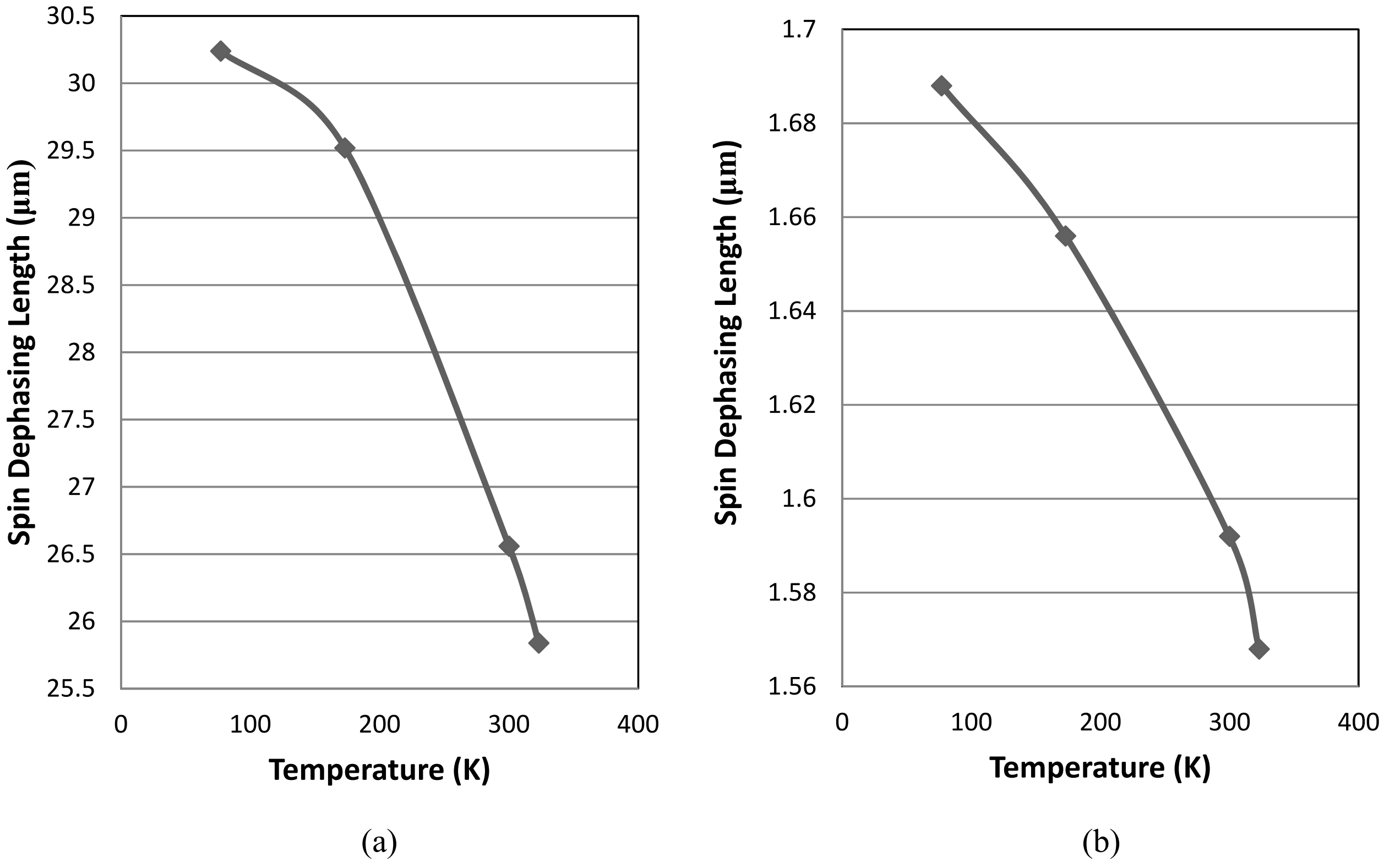

Fig.10. Variation of spin dephasing length with temperature for GaAs (a) nanowire (b) 2-D channel

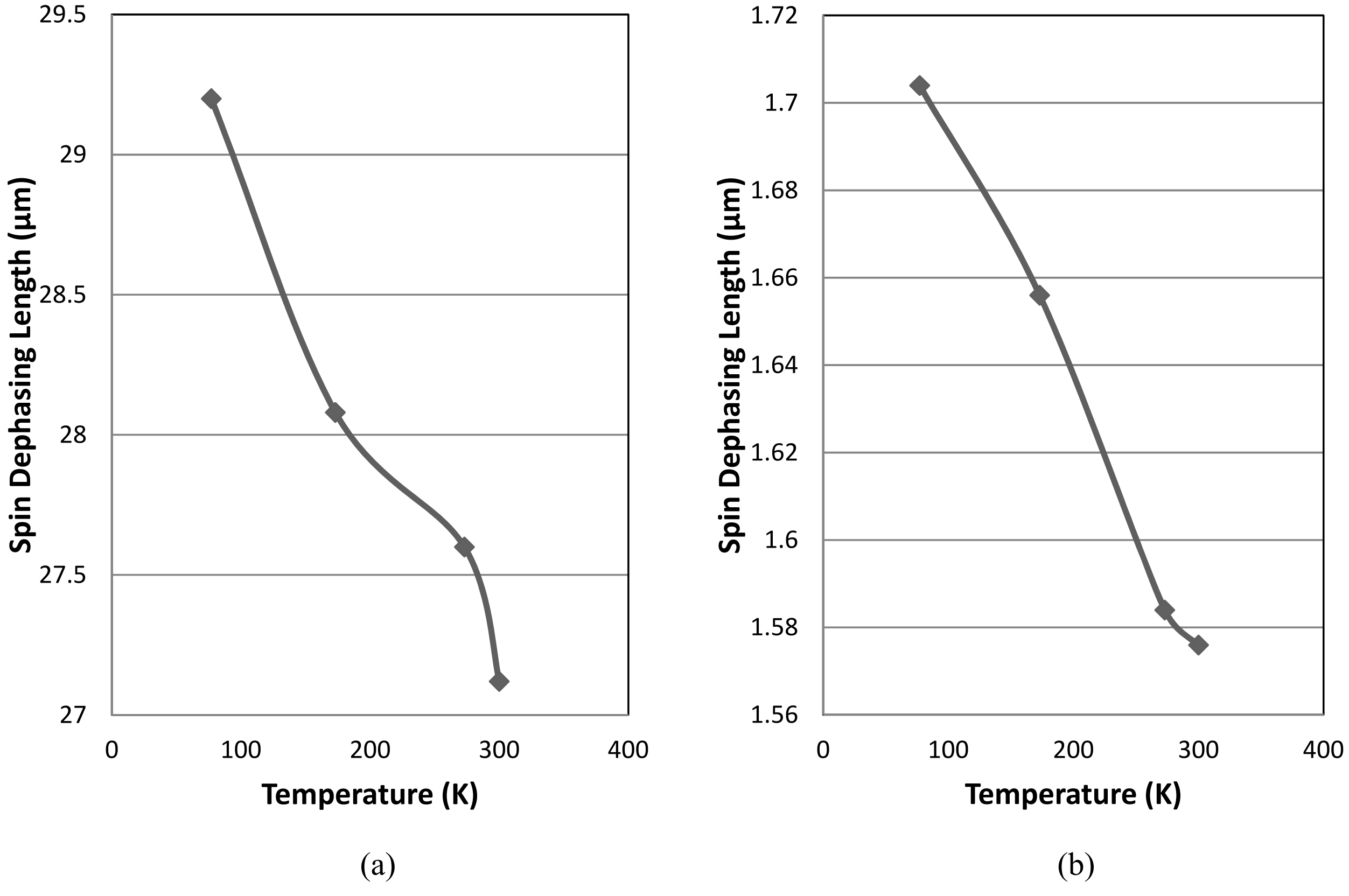

Fig.11. Variation of spin dephasing length with temperature for GaP (a) nanowire (b) 2-D channel

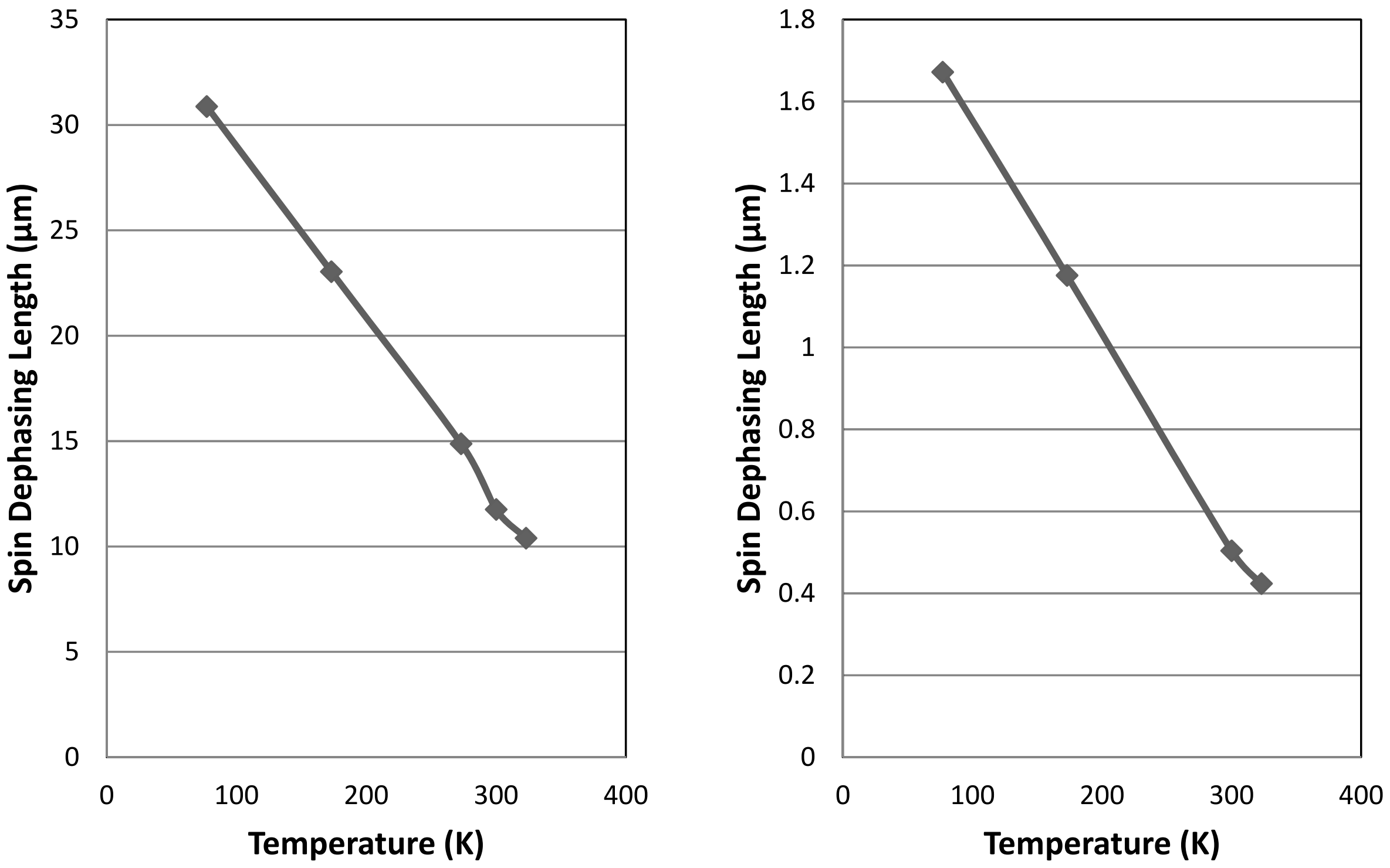

(a) (b)

Fig.12. Variation of spin dephasing length with temperature for GaSb (a) nanowire (b) 2-D channel

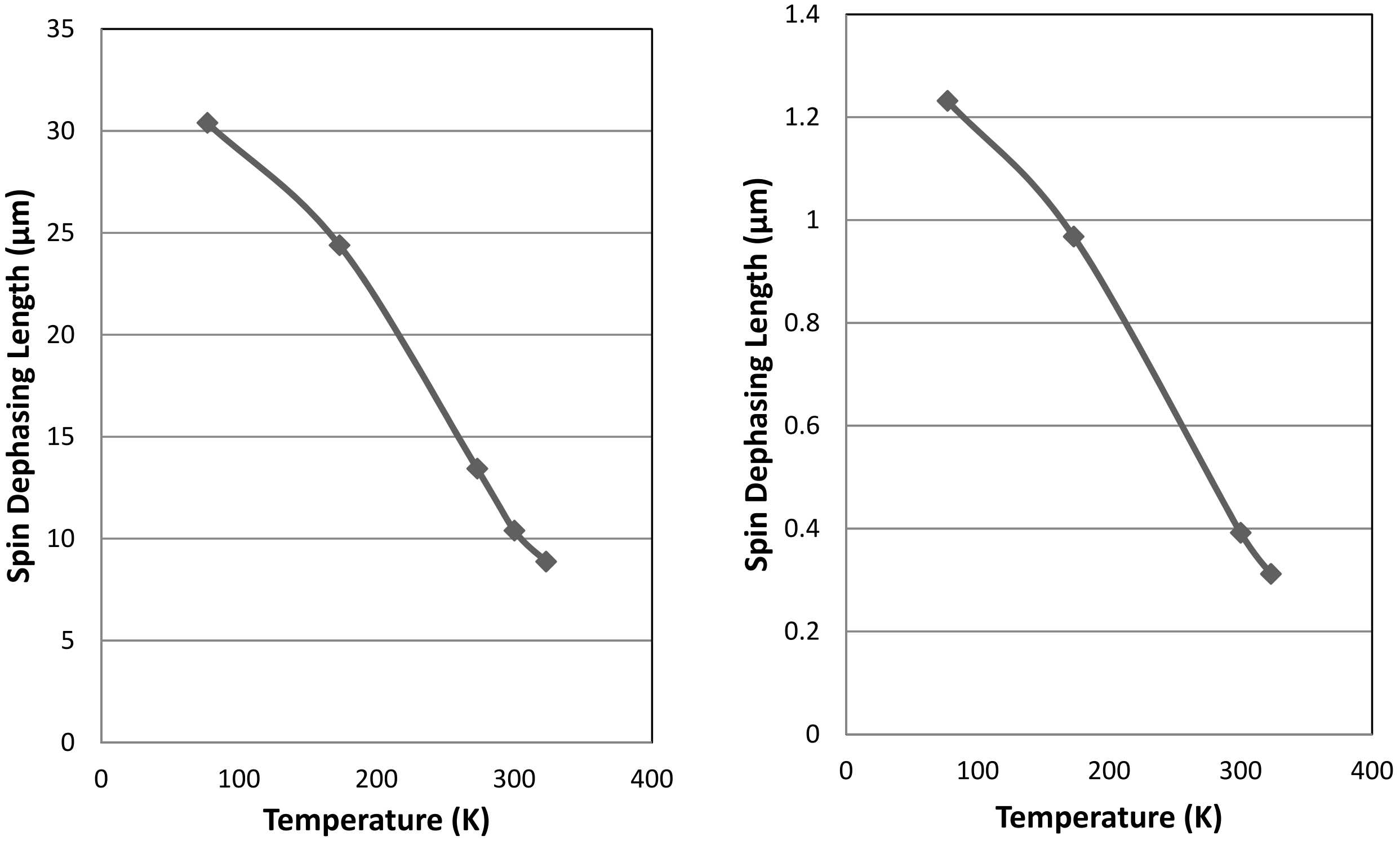

(a) (b)

Fig.13. Variation of spin dephasing length with temperature for InAs (a) nanowire (b) 2-D channel

## Conclusion

We have investigated the spin transport properties of III-V semiconductors for 1-D and 2-D channel. Simulation is performed for GaAs, GaP, GaSb and InAs to compare their spin properties for 1-D and 2-D channel. We have also analyzed the variation of spin dephasing length with driving electric field and temperature. The pairs of materials, namely, (GaAs, GaP) and (GaSb, InAs) have approximately same dephasing length where as large difference of length is observed among these two pairs of materials. With increase in temperature, the spin dephasing length decreases which means the device can perform better at lower temperature. At higher driving electric field, spin dephasing length increases with increase in driving electric field whereas at lower values of driving electric field the dependence is nonmonotonic.